\documentclass[aps,prb,preprint,superscriptaddress,amsmath,amssymb]{revtex4}

\newcommand {\CA}{Cd$_{3}$As$_{2}$}
\newcommand {\CAS}{Cd$_{3}$(As$_{1-y}$Sb$_{y}$)$_{2}$}
\newcommand {\Rxx}{$R_\mathrm{xx}$}
\newcommand {\Ryx}{$R_\mathrm{yx}$}

\usepackage{graphicx}
\usepackage{dcolumn}
\usepackage{bm}
\usepackage{color}
\usepackage{ulem}

\begin{document}


\title{Enhancement of spin-orbit coupling in Dirac semimetal {\CA} films by Sb-doping}



\author{Yusuke Nakazawa}
\affiliation{Department of Applied Physics and Quantum-Phase Electronics Center (QPEC), the University of Tokyo, Tokyo 113-8656, Japan}

\author{Masaki Uchida}
\email[Author to whom correspondence should be addressed : ]{m.uchida@phys.titech.ac.jp}
\affiliation{Department of Applied Physics and Quantum-Phase Electronics Center (QPEC), the University of Tokyo, Tokyo 113-8656, Japan}
\affiliation{PRESTO, Japan Science and Technology Agency (JST), Tokyo 102-0076, Japan}

\author{Shinichi Nishihaya}
\affiliation{Department of Applied Physics and Quantum-Phase Electronics Center (QPEC), the University of Tokyo, Tokyo 113-8656, Japan}

\author{Mizuki Ohno}
\affiliation{Department of Applied Physics and Quantum-Phase Electronics Center (QPEC), the University of Tokyo, Tokyo 113-8656, Japan}

\author{Shin Sato}
\affiliation{Department of Applied Physics and Quantum-Phase Electronics Center (QPEC), the University of Tokyo, Tokyo 113-8656, Japan}

\author{Masashi Kawasaki}
\affiliation{Department of Applied Physics and Quantum-Phase Electronics Center (QPEC), the University of Tokyo, Tokyo 113-8656, Japan}
\affiliation{RIKEN Center for Emergent Matter Science (CEMS), Wako 351-0198, Japan}


\date{\today}

\begin{abstract}
We present a study on magnetotransport in films of the topological Dirac semimetal {\CA} doped with Sb grown by molecular beam epitaxy.
In our weak antilocalization analysis, we find a significant enhancement of the spin-orbit scattering rate, indicating that Sb doping leads to a strong increase of the pristine band-inversion energy.
We discuss possible origins of this large enhancement by comparing Sb-doped {\CA} with other compound semiconductors.
Sb-doped {\CA} will be a suitable system for further investigations and functionalization of topological Dirac semimetals.
\end{abstract}


\maketitle




Topologically non-trivial phases are commonly characterized by band inversion induced by large spin-orbit coupling \cite{RMP2010Hasan,RMP2011Qi, RMP2018Armitage}.
As seen in the field of topological insulators, tuning the band inversion through modulation of the spin-orbit coupling has advanced the understanding such as of topological phase transitions \cite{PRB2011Guo, Science2011Xu}.
A topological Dirac semimetal, which is characterized by stable pairs of bulk Dirac dispersions in the presence of band inversion, is regarded as an ideal platform to study a variety of topological phases \cite{PRL2012Young, PRB2012Wang, PRB2013Wang, NatCom2014Yang}.
In particular, tuning the band inversion in topological Dirac semimetals does not only serve as a source of various topological phase transitions but also allows to control the separation of the Dirac points in the momentum space, leading to exotic quantum transport phenomena.
Moreover, the quantum spin Hall insulator phase in quantum confinement \cite{PRB2013Wang} can be stabilized by increasing the band-inversion energy.
These few examples already suggest that gaining control over the tunability of the spin-orbit coupling is strongly desired for investigations of topological Dirac semimetals.

{\CA} is prototypical topological Dirac semimetal \cite{NatMat2014Liu, NatMat2014Jeon, Nature2016Moll, NatCom2017Uchida, NatCom2017Zhang, PRL2019Lin}, exhibiting a simple band structure, where only Cd 5$s$ and As 4$p$ bands, which form the Dirac dispersions, exist over a wide energy range around the Fermi level, as shown in Fig. 1(a).
As an example of exotic quantum transport, quantum Hall states mediated by both the surface and bulk states \cite{Nature2019Zhang, NatCom2019Nishihaya} have been reported in this compound.
Although chemical doping has been considered effective for tuning the band inversion as in the cases of topological insulators, it has been limited to Zn- and P-doping because of the high volatility of {\CA} itself \cite{SciRep2017Lu, PRB2018Thirupathaiah}.
Also in high-crystallinity {\CA} films obtained by solid-phase epitaxy \cite{NatCom2017Uchida, NatCom2019Nishihaya, SciRep2018Nakazawa, PRB2019Uchida}, Zn-doping has been demonstrated to successfully modulate the quantum transport and to induce a topological phase transition to a trivial insulator simply by reducing the spin-orbit coupling \cite{NatCom2019Nishihaya, PRB2018Nishihaya, SciAdv2018Nishihaya}.
Unfortunately, however, solid-phase epitaxy does not work on other dopant elements, and so far, chemical doping efficiently enhancing the spin-orbit coupling has not been reported for {\CA}.

Doping Sb which exhibiting a large spin-orbit coupling is promising to overcome this issue.
From the band-gap change measured in P-doped {\CA}, 10 \% Sb-doping is expected to increase the valence-band-maximum energy by 80 meV, judged from the band alignment of conventional compound semiconductors \cite{PRB2018Thirupathaiah, PRB2014Hinuma}.
In addition to enhancement of the band-inversion energy, lowering of the Fermi level is also expected, as shown in Fig. 1(a).
This is because the main origin of the $n$-type carriers in {\CA} is the deficiency at the anion site (As and Sb), and since Sb is less volatile than As, a decrease of the carrier density can be expected.
Considering that it is difficult to dope especially volatile atoms into {\CA} under thermal equilibrium, molecular beam epitaxy, which is thermally a non-equilibrium process as compared to solid-phase epitaxy, is expected more suitable to dope Sb into {\CA} effectively.
For undoped {\CA}, high-quality films have been grown also by molecular-beam epitaxy, and a quantum Hall state has been observed for the films with two-dimensional Fermi surface \cite{PRL2018Schumann}.
Quantum Hall states have also been observed for the films with thickness of three-dimensional Fermi surface, where involved magnetic orbits can be described neither by a simple two-dimensional nor by a three-dimensional electronic state \cite{APLMat2019Nakazawa}.
In this letter, we report transport signatures of an enhancement of spin-orbit coupling in single-crystalline {\CAS} films grown by molecular-beam epitaxy.
A systematic analysis of weak antilocalization in films with various Sb concentrations reveals that the spin-orbit scattering rate is dramatically enhanced by Sb doping.

{\CAS} films with $y$ ranging from 0.005 to 0.24 were grown by molecular-beam epitaxy, following the same procedure described in Ref. \cite{APLMat2019Nakazawa}.
The single-crystalline (1 1 1)A CdTe substrate was heated at 160 $^{\circ}$C during the growth.
The molecular beams were simultaneously provided from conventional Knudsen cells containing 6N Cd and 6N Sb and a valved cracker source containing 7N5 As.
The beam equivalent pressures were set to 1$\times$10$^{-4}$ Pa for Cd and As$_{4}$ and 0.05 - 5$\times$10$^{-5}$ Pa for Sb.
Transport measurements were performed by using the conventional four-terminal method on film samples of 5 mm $\times$ 1 mm size.

Figure 1(b) shows a cross-sectional scanning transmission electron microscopy (STEM) image taken on a {\CAS} film with $y$ = 0.15.
{\CAS} layers are clearly formed from the heterointerface to the CdTe substrate.
An x-ray diffraction (XRD) $\theta$-2$\theta$ scan for the $y$ = 0.15 film in Fig. 1(c) shows the reflections from the \{1 1 2\} lattice planes without any impurity phases.
As shown in Fig. 1(d), XRD scans show a systematic peak shift with increasing $y$, indicating that the lattice constant systematically increases upon Sb doping.
The film thicknesses determined by atomic force microscopy are 186 nm, 156 nm, 82 nm, 52 nm, and 83 nm for $y$ = 0.005, 0.07, 0.12, 0.15, and 0.24 samples, respectively.
These thickness variation is caused by difficulty of estimating accurate growth rate for each sample due to Sb doping effects.
However, transport properties of the {\CAS} samples show systematic dependence on the Sb concentration as presented below, indicating that the thickness variation does not have a considerable effect on discussions.

Figure 2(a) shows longitudinal resistance {\Rxx} and Hall resistance {\Ryx}, taken on the $y$ = 0.15 film with sweeping the magnetic field at 2 K.
{\Ryx} has a positive slope and is slightly bent at low fields.
This indicates that the conduction is dominated by holes ($p$ $\sim$ 10$^{18}$ cm$^{-3}$) and that there is a small number of remnant electrons ($n$ $\sim$ 10$^{15}$ cm$^{-3}$).
The low-temperature carrier mobility and carrier density obtained for all the {\CAS} films are summarized in Figs. 2(b) and 2(c).
As also confirmed from the Hall resistance in Figs. 2(d) - 2(i), the electron carrier density is systematically reduced with Sb doping.
For $y = 0.12$, the Fermi energy measured from the Dirac points is estimated from its carrier density to be about 40 meV, and the saddle point energy is estimated to be about 100 meV \cite{PRB2018Thirupathaiah, PRB2014Hinuma}, suggesting that the Fermi energy is located inside the band-inversion region.
At $y$ = 0.15, the hole carriers become dominant.
This indicates that the deficiency at As sites is greatly suppressed by Sb doping as expected, and eventually the system becomes $p$ type.
As commonly observed in III-V and II-VI compound semiconductors, the valence band level is higher in heavier anions \cite{PRB2014Hinuma, Nature2003VanDeWalle}.
Based on these observations in the conventional compound semiconductors, the valence band level of {\CA} is also expected to be shifted higher by Sb doping, and hole carriers are considered to become easier to be induced as a result of the shallower valence band level.

Magnetoresistivity ratio $\mathrm{MRR} \equiv [\rho(B)-\rho(0)] / \rho(0) = \Delta \rho(B)/\rho(0)$ of the {\CAS} films at 2 K are shown in Figs. 2(j) - 2(n).
Due to Sb doping, the shape of the MRR curves changes from parabolic-like in the lightly doped region to cusp-like in the heavily-doped region.
This change indicates that the Sb doping suppresses the classical magnetoresistivity and enhances weak antilocalization effects as discussed below.
The magnitude of the classical magnetoresistivity originating from the Lorentz force is expressed by 
\begin{equation}
  \frac{\Delta \rho_\mathrm{orb}(B)}{\rho(0)} \approx (\mu B)^\mathrm{2},
  \label{eq:MR_orb}
\end{equation}
with the carrier mobility $\mu$ \cite{MinM1989Pippard}, and, therefore, its contribution decreases when the carrier mobility is reduced.

The observed magnetoresistivity $\rho_\mathrm{total}$ can be expressed as $\rho_\mathrm{total} = \rho_\mathrm{orb} + \rho_\mathrm{WAL}$, where $\rho_\mathrm{orb}$ is the classical magnetoresistivity and $\rho_\mathrm{WAL}$ is a quantum correction to the magnetoresistivity due to weak antilocalization.
According to Kohler's rule \cite{PhysC2002Luo}, Eq. (\ref{eq:MR_orb}) can be rewritten as
\begin{equation}
  \frac{\Delta\rho_\mathrm{orb}(B)}{\rho(0)} =  k\left(\frac{B}{\rho(0)}\right)^{2}
  \label{eq:Kohler_orb}
\end{equation}\\
with a temperature-independent constant $k$, as long as the carrier density is independent of temperature.
In the present case, $k$ can be obtained by fitting the total magnetoresistivity at high temperatures ($T$ $\sim$ 50 K), above which only $\Delta\rho_\mathrm{orb}(B)$ is observed.
By using this $k$ value, $\Delta\rho_\mathrm{WAL}(B)$ at low temperatures can be derived by subtracting $\Delta\rho_\mathrm{orb}(B)$ from $\Delta\rho_\mathrm{total}(B)$.
Details about the subtraction of the classical magnetoresistivity can be found in the Supplementary Materials \cite{Supple}.

Figure 3(a) shows $\Delta\rho_{\mathrm{WAL}}(B) \equiv \Delta\rho_{\mathrm{total}}(B) - \Delta\rho_{\mathrm{orb}}(B)$ at 2 K for all examined {\CAS} films.
Before moving on to a detailed discussion of the weak antilocalization, we first check the dimensionality of the {\CAS} films.
The dimensionality of weak localization and weak antilocalization is determined by comparing the film thickness $t$ to the dephasing length $L_\phi = \sqrt{D\tau_\phi}$, with the diffusion coefficient $D$ and the dephasing scattering time $\tau_\phi$.
The dephasing process is dominated by inelastic scatterings including electron-phonon and electron-electron scattering \cite{JPCM2002Lin}.
When $L_\phi$ is smaller than $t$, the quantum interference process causing the weak antilocalization is treated as three-dimensional (Fig. 3(b)), and otherwise, as two-dimensional (Fig. 3(c)).
In the case of our {\CAS} films, we employed both two-dimensional and three-dimensional models to obtain $L_\phi$ and compare it with $t$ to determine the dimensionality of the weak antilocalization.
For $y = 0.07 - 0.24$, the experimental data can be well fitted by the three-dimensional weak antilocalization model \cite{JPCM2002Lin, JPSJ1981Fukuyama, JPSJ1981Maekawa, JPhys1989Baxter} as presented in Figs. 3(d) - 3(g), and $L_\phi$ obtained from this model is smaller than $t$.
This suggests that the weak antilocalization observed in these films can be treated as three-dimensional.
On the other hand, for $y = 0.005$, a narrow dip observed at low fields ($B < 40$ mT) cannot be described by the three-dimensional model.
Instead, it is well fitted by the two-dimensional weak antilocalization model \cite{PTP1980Hikami, PhysRep1984Bergmann}, and $L_\phi$ for the $y = 0.005$ film obtained from the two-dimensional model satisfies $L_\phi > t$, suggesting that the weak antilocalization observed in this film is of two-dimensional nature.
Two-dimensional weak antilocalization has also been reported for undoped {\CA} films previously \cite{SciRepZhao2016}.
In the following discussion, we focus on the three-dimensional weak antilocalization observed in $y = 0.07 - 0.24$ films.
See also the Supplementary Materials for further discussion on the dimensionality and the fitting of the $y = 0.005$ film \cite{Supple}.

Three-dimensional weak localization and weak antilocalization including the effect of the spin-orbit coupling are expressed as \cite{JPCM2002Lin, JPSJ1981Fukuyama, JPSJ1981Maekawa, JPhys1989Baxter, approx}
\begin{widetext}
  \begin{eqnarray}
  \frac{\Delta\rho_\mathrm{WAL}(B)}{\rho^{2}(0)} &=&
  \frac{e^{2}}{2\pi^{2}\hbar}
  \sqrt{\frac{eB}{\hbar}}
  \Biggl(
  \frac{1}{2\sqrt{1-\gamma}}
  \biggl[
    f_{3}\left(
      \frac{B}{B_{-}}
    \right)
    -f_{3}\left(
      \frac{B}{B_{+}}
    \right)
  \biggr]
  - f_{3}\left(
      \frac{B}{B_{2}}
    \right) \nonumber \\
  &-& \sqrt{\frac{4B_\mathrm{SO}}{3B}}
  \left[
    \frac{1}{\sqrt{1-\gamma}}
    \left(
     \sqrt{t_{+}}
     - \sqrt{t_{-}}
    \right)
    +\sqrt{t}
    -\sqrt{t+1}
  \right]
  \Biggr),
   \label{eq:3DWAL}
  \end{eqnarray}
\end{widetext}
where
\begin{align*}
  &\gamma = 
  \left[
  \frac{3g^{*}\mu_{B}B}{4eD(2B_{\mathrm{SO}} - B_{0})}
  \right]^{2},\\
  &B_{\pm} = B_{\phi} + \frac{1}{3} \left( 2B_{\mathrm{SO}} - B_{\mathrm{0}} \right) \left( 1 \pm \sqrt{1-\gamma} \right),\\
  &B_{\phi} = B_{i} + B_{\mathrm{0}},\;\;\;
  B_{2} = B_{i} + \frac{1}{3}B_{\mathrm{0}} + \frac{4}{3}B_{\mathrm{SO}},\\
  &t = \frac{3B_{\phi}}{2\left( B_{\mathrm{SO}} - B_{0} \right)},\;\;\;
  t_{\pm} = t + \frac{1}{2}\left( 1 \pm \sqrt{1-\gamma} \right),\\
  &f_{3}(z) =
  \sum_{N=0}^{\infty}
  \left[
  2\left(
  \sqrt{N + 1 + z} - \sqrt{N + z}
  \right)
  - \frac{1}{\sqrt{N + \frac{1}{2} + z}}
  \right].
\end{align*}
Here, $g^{\ast}$ is the Land\'{e} factor, $\mu_{\mathrm{B}}$ is the Bohr magneton, and $D = \mu k_{\mathrm{B}}T/e$ is the diffusion constant.
The characteristic fields $B_{j}$ ($j$ = $i$, SO, and 0) are related to the characteristic scattering times through $B_{\mathrm{j}} = \hbar/4eD\tau_{\mathrm{j}}$, where the indices denote inelastic, spin-orbit, and residual scattering, respectively.
The dephasing scattering time $\tau_{\phi}$ is defined by
\begin{equation}
  \frac{1}{\tau_{\phi}(T)} =
  \frac{1}{\tau_{0}}
  + \frac{1}{\tau_{i}(T)},
\end{equation}
where its temperature dependence is governed by $\tau_{i}$, and its low-temperature limit is determined by temperature-independent $\tau_{0}$ \cite{JPCM2002Lin}.
In the spin-orbit scattering process, on the other hand, not only the momentum but also the spin direction of the scattered electrons is changed due to the spin-orbit interaction \cite{JETP1962Abrikosov}.
In the limit of $1/\tau_{i} \gg 1/\tau_{\mathrm{SO}}$, Eq. (\ref{eq:3DWAL}) describes weak localization, and in the opposite limit, $1/\tau_{\mathrm{SO}} \gg 1/\tau_{i}$, it describes weak antilocalization.

Temperature dependences of $1/\tau_{\phi}$ and $1/\tau_{\mathrm{SO}}$ are shown in Figs. 4(a) and 4(b).
An exponential temperature dependence of $1/\tau_{\phi}$ is confirmed for all films.
The extracted exponents $p = 1.9 - 3.1$ are quite typical for three-dimensional systems in the diffusive regime.
Similar exponents $p \sim 2 - 4$ have been previously reported for conventional compound semiconductors \cite{JPCM2002Lin, PRB2002Oszwaldowski, PRB2009Peres}.

$1/\tau_{\mathrm{SO}}$ is significantly enhanced by Sb doping, as shown in Fig. 4(b).
The spin-orbit scattering cross section of an impurity $\sigma_{\mathrm{SO}}$ has been previously studied theoretically and experimentally \cite{PRL1992Geier, PRL1992Papanikolaou, JdePhys1982Monod, PRB2009Fedorov}, and it is expressed as
\begin{equation}\label{eq:Sso}
  \sigma_{\mathrm{SO}} \sim
  \frac{1}{\tau_{\mathrm{SO}}} \sim
  \sum_{l\geqq 0} \frac{l(l+1)}{2l+1} \xi_{l}^{2} n_{l}^{2}(E_{F}),
\end{equation}
where $l$ is the azimuthal quantum number for impurity atoms, $\xi_{l}$ is half of the spin-orbit constant \cite{PRL1992Papanikolaou}, and $n_{l}(E_{F})$ is the angular momentum resolved local density of states of the impurity atoms at the Fermi level.
Eq. (\ref{eq:Sso}) suggests that $1/\tau_{\mathrm{SO}}$ largely depends on the local density of $p$ states ($l=1$) of impurity at the Fermi level ($n_{l=1}(E_{F})$) \cite{PRL1992Papanikolaou}, and, therefore, the observed enhancement of $1/\tau_{\mathrm{SO}}$ in {\CAS} is considered a result of a larger $n_{l=1}(E_{F})$ with an increased impurity density by Sb doping.

The relation between $1/\tau_{\mathrm{SO}}$ and $1/\tau_{\phi}$ obtained for the {\CAS} films is summarized in Fig. 4(c), and compared with previously reported data on both three-dimensional and two-dimensional systems \cite{PRB2002Oszwaldowski, PRB2009Peres, PRB2003Studenikin, PRB2008Grbic, PRB2009Kallaher, PRB2006Thillosen, APL2006Thillosen}.
In particular, a large enhancement of $1/\tau_{\mathrm{SO}}$ has been observed for the $p$-type {\CAS} films, and this may suggest that a similar enhancement can be also expected for conventional compound semiconductors where the following two conditions are satisfied: (i) the doping takes place at the anion sites and (ii) the Fermi level is tuned into the valence band.
For the compound semiconductors shown in Fig. 4(c), at least one of these conditions is not satisfied.
It is further suggested that a large enhancement of $1/\tau_{\mathrm{SO}}$ can be expected, for example, for $p$-type GaAs$_{1-x}$Sb$_{x}$ and $p$-type InP$_{1-x}$As$_{x}$, which satisfy the aforementioned two conditions but these samples have not been examined yet.

In summary, we have systematically investigated magnetotransport of {\CAS} films with varying Sb concentrations $y = 0.005 - 0.24$.
Weak antilocalization is commonly observed in the {\CAS} films and has been analyzed using a three-dimensional model.
In the analysis, $1/\tau_{\mathrm{SO}}$ is greatly enhanced upon increasing the Sb concentration.
In particular, the Fermi energy of the $y=0.12$ film is estimated to be located inside the band-inversion region, based on the band alignment of conventional compound semiconductors.
Our demonstration of a large enhancement of the spin-orbit coupling in {\CAS} films paves the way for further investigations and functionalization of topological Dirac semimetals, including the realization of a robust quantum spin Hall insulator phase under confinement conditions as in HgTe/CdTe quantum wells \cite{PRB2013Wang, Science2006Bernevig, Science2007Konig}.
Tuning a confinement-induced sub-band position and the Fermi energy inside the band-inversion region is considered more feasible with the increased band inversion in {\CAS}.

We thank N. Nagaosa and M. Tokunaga for fruitful discussions.
We also thank M. Kriener for proofreading of the manuscript.
This work was supported by JST PRESTO Grant No. JPMJPR18L2 and CREST Grant No. JPMJCR16F1, Japan, and by Grant-in-Aids for Scientific Research (B) No. JP18H01866 from MEXT, Japan.

\newpage
\begin{figure}
  \includegraphics[width = 13 cm]{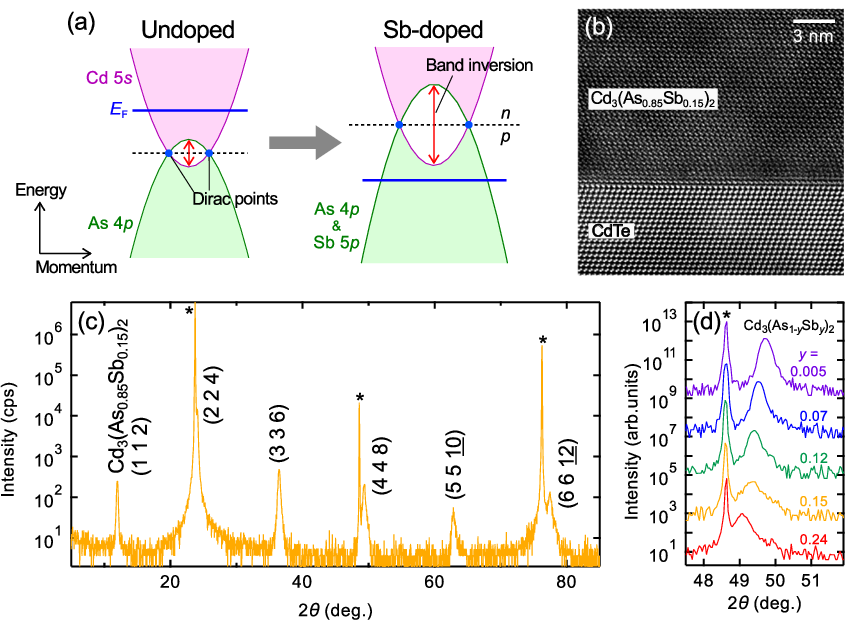}
  \caption{\label{fig:Fig1}
  (a) Schematic band-structure change induced by Sb doping in {\CA}.
  The band inversion is expected to be enhanced by Sb doping which exhibits a larger spin-orbit coupling than As.
  Electron carriers caused by As and Sb deficiency are also expected to be reduced.
  (b) Cross-section HAADF-STEM image and (c) XRD $\theta - 2\theta$ scan of a Cd$_{3}$(As$_{0.85}$Sb$_{0.15}$)$_{2}$ film grown on a CdTe substrate.
  The asterisks denote peaks from the CdTe substrate.
  (d) XRD $\theta - 2\theta$ scans magnified around the (4 4 8) {\CAS} film peak with different Sb concentrations $y$.
  The Sb concentrations are estimated using Vegard’s law, which linearly interpolates the lattice constants between $y$ = 0 and 0.15 films determined by EDX analyses.
  }
\end{figure}

\newpage
\begin{figure}
  \includegraphics[width = 13 cm]{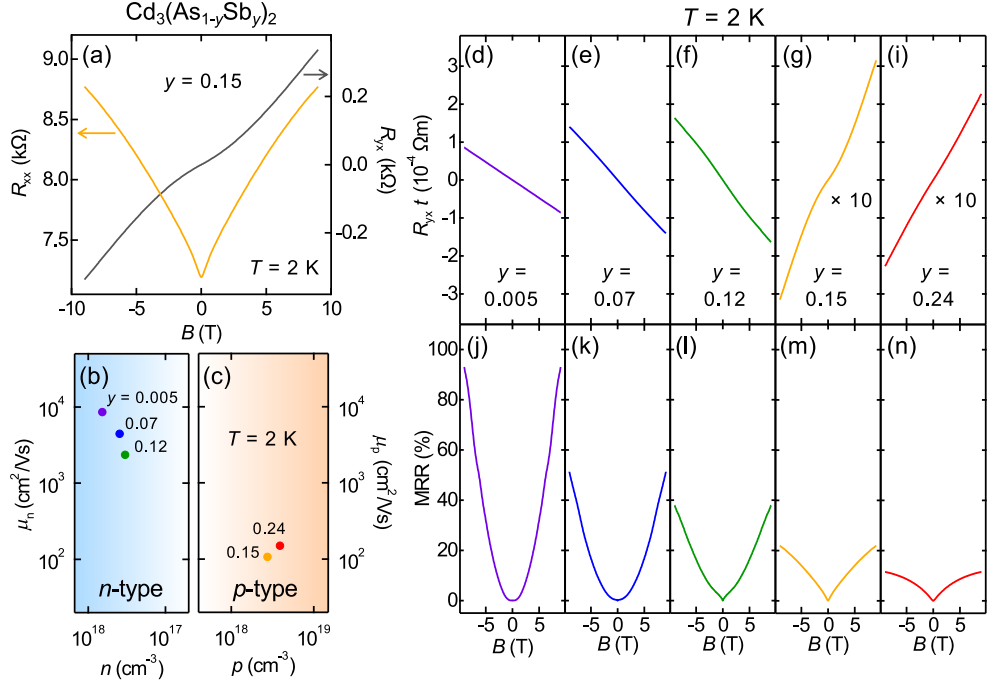}%
  \caption{\label{fig:Fig2}
  (a) Longitudinal resistance {\Rxx} and Hall resistance {\Ryx} of Cd$_{3}$(As$_{0.85}$Sb$_{0.15}$)$_{2}$ film at 2 K. The magnetic field $B$ is applied out of plane.
  Mobility vs carrier density summarized for our {\CAS} films in (b) $n$-type and (c) $p$-type regions.
  (d)-(i) Hall resistance times thickness {\Ryx}$t$ and (j)-(n) magnetoresistance ratio for our {\CAS} films at 2 K.
  In (g) and (i), the curves are multiplied by ten for clarity.
  }
\end{figure}

\newpage
\begin{figure}[b]
  \includegraphics[width = 13 cm]{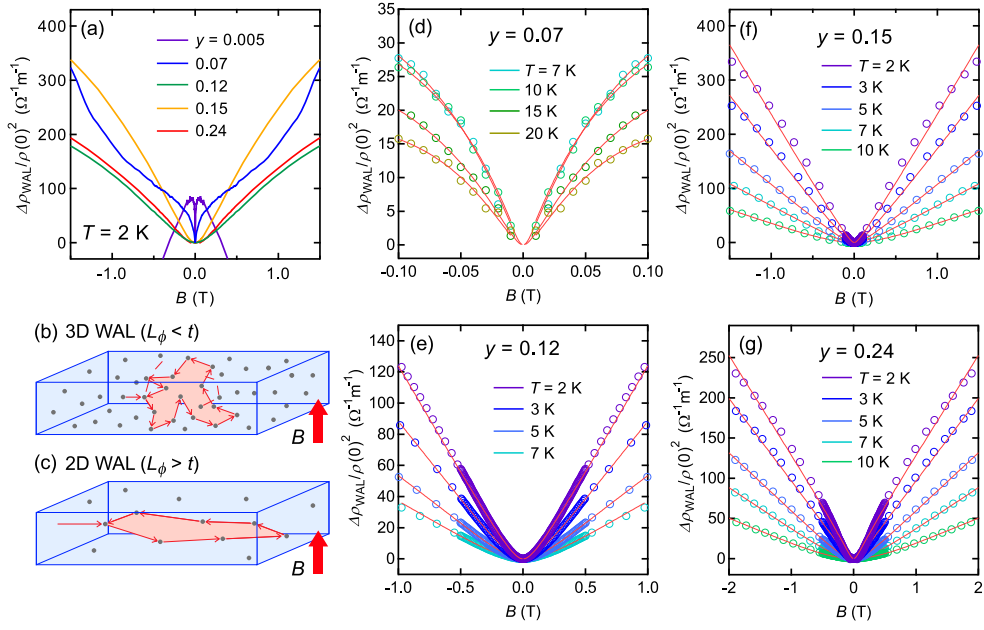}%
  \caption{\label{fig:Fig3}
  (a) Normalized magnetoresistivity $\Delta \rho_{\mathrm{WAL}} / \rho^{2}(0) = [\Delta \rho_{\mathrm{total}} - \Delta \rho_{\mathrm{orb}}] / \rho_{\mathrm{total}}^{2} (0)$ of the {\CAS} films at 2 K.
  Trajectories of the scattered electrons for (b) three-dimensional and (c) two-dimensional weak antilocalization.
  (d)-(g) Temperature dependence of the normalized magnetoresistivity for $y$ = 0.07, 0.12, 0.15, and 0.24.
  The red curves are fits to the three-dimensional weak antilocalization model (Eq. (3)).
  The theory assumes that the magnetic field is within the range satisfying $\gamma < 1$ (see Eq. (\ref{eq:3DWAL}) in main text).
  The magnetoresistivity below 7 K is not shown for $y = 0.07$ because $L_{\phi} < t$ is not satisfied at lower temperature (see also the Supplementary Material).
}
\end{figure}

\newpage
\begin{figure}
  \includegraphics[width = 13 cm]{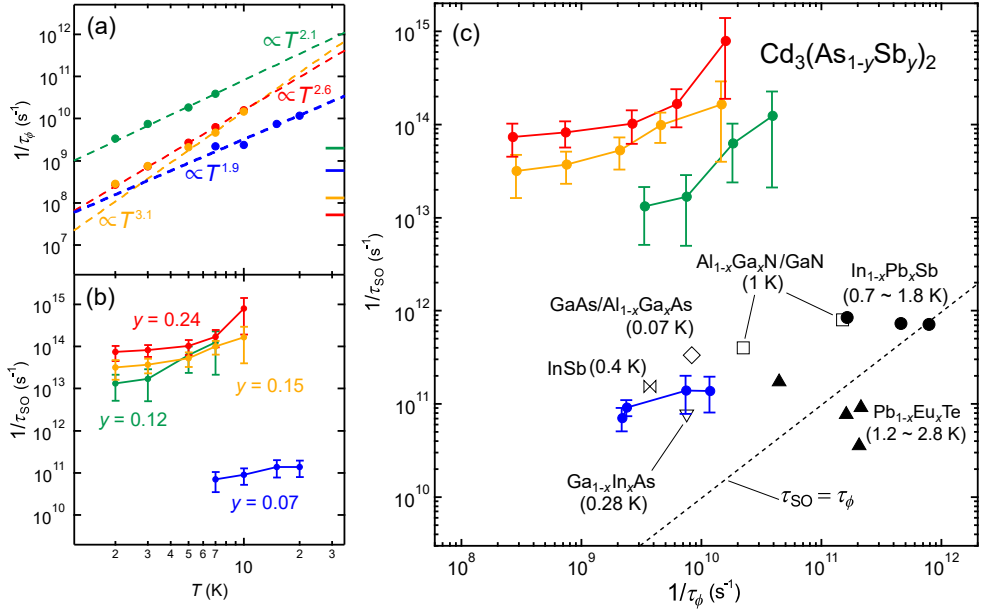}%
  \caption{\label{fig:Fig4}
  (a) Dephasing scattering rate $1/\tau_{\mathrm{\phi}}$ and (b) spin-orbit scattering rate $1/\tau_{\mathrm{SO}}$ as a function of temperature for $y$ = 0.07, 0.12, 0.15, and 0.24.
  The horizontal bars on the right axis in (a) represent the temperature-independent residual scattering rate extracted from the fit at 7 K for $y$ = 0.07 and at 2 K for $y$ = 0.12, 0.15, and 0.24.
  (c) Spin-orbit scattering rate $1/\tau_{\mathrm{SO}}$ vs dephasing scattering rate $1/\tau_{\phi}$ summarized for the {\CAS} films. 
  Data taken at different temperatures as shown in (a) and (b) are plotted.
  For comparison, $1/\tau_{\mathrm{SO}}$ vs $1/\tau_{\phi}$ previously reported for three-dimensional systems ($\bullet$ In$_{1-x}$Pb$_x$Sb \cite{PRB2002Oszwaldowski}, $\blacktriangle$ Pb$_{1-x}$Eu$_x$Te \cite{PRB2009Peres}) 
  and two-dimensional systems ($\bigtriangledown$ Ga$_{1-x}$In$_x$As \cite{PRB2003Studenikin}, $\diamond$ GaAs/Al$_{1-x}$Ga$_x$As \cite{PRB2008Grbic}, $\bowtie$ InSb \cite{PRB2009Kallaher}, $\Box$ Al$_{1-x}$Ga$_x$N/GaN \cite{PRB2006Thillosen, APL2006Thillosen}) are also plotted.
  $1/\tau_{i}$ ($1/\tau_{\phi}$ at $\tau_{0} \rightarrow \infty$) is shown for Pb$_{1-x}$Eu$_x$Te instead of $1/\tau_{\phi}$.
}
\end{figure}


\begin{thebibliography}{100}
\bibitem{approx} The summation function $f_{3}$ can be approximated by the formula\\
$f_{3}(z)\approx 2\left(
 \sqrt{2 + \frac{1}{z}} - \sqrt{\frac{1}{z}}
\right)
- \left[
 \left(
   \frac{1}{2} + \frac{1}{z}
 \right)^{- \frac{1}{2}}
 + \left(
   \frac{3}{2} + \frac{1}{z}
 \right)^{- \frac{1}{2}}
\right]
+ \frac{1}{48}
\left(
 2.03 + \frac{1}{z}
\right)^{- \frac{3}{2}}$\\
with sufficient accuracy. We used this approximation in the present analysis.
\end{thebibliography}
\end{document}